\documentclass[aps,pre,twocolumn,showpacs,nofootinbib]{revtex4-1}
\usepackage{graphicx}
\usepackage{subfigure}
\usepackage{url}
\usepackage{hyperref}
\usepackage{inconsolata}
\usepackage{amsmath}
\usepackage[capitalize]{cleveref}
\usepackage{braket}

\begin{document}
\title{Field-induced freezing in the unfrustrated Ising antiferromagnet}

\author{Adam Iaizzi}
\email{iaizzi@bu.edu}
\affiliation{Department of Physics, National Taiwan University, No. 1, Section 4, Roosevelt Road, Taipei 10607, Taiwan}

\date{\today}
\begin{abstract}
We study instantaneous quenches from infinite temperature to well below $T_c$ in the two-dimensional square lattice Ising antiferromagnet in the presence of a longitudinal external magnetic field. 
Under single-spin-flip Metropolis algorithm Monte Carlo dynamics, this protocol produces a pair of magnetization plateaus that prevent the system from reaching the equilibrium ground state except for some special values of the field. 
We explain the plateaus in terms of local spin configurations that are stable under the dynamics. 
\end{abstract}

\maketitle

\section{Introduction}

The Ising model is the basis for much of our modern understanding of both magnetism and phase transitions and serves as the go-to proving ground for new theoretical and numerical methods. 
After a century of intensive study, this deceptively simple model has yet to reveal all its secrets. 
One area of ongoing research is the Ising model's dynamics, which provide a rich variety of behaviors that can be generalized to understand more complicated systems. 
We are interested in instantaneous quenches to very low temperatures (well below $T_c$). 
In 2001, Spirin \textit{et al.} showed that in the square-lattice Ising ferromagnet, Monte Carlo dynamics cannot always reach ground state after a quench to zero temperature \cite{spirin2001t0,spirin2001}. 
Instead, the system can become permanently stuck in states with stable stripe defects. 
Interest in this phenomenon increased when the same group proved \cite{barros2009} that the probability of becoming stuck is related to critical percolation theory.  
This freezing and variations on it have been the subject of a steady stream of research \cite{olejarz2012,blanchard2013,mullick2017,godreche2018,denholm2019,denholm2020} along with Ising model dynamics more broadly \cite{blanchard2014,cugliandolo2016,blanchard2017,corberi2017,tartaglia2018,lopes2020}. 

In this paper, we take a different approach: rather than introducing disorder or new interactions, we consider the Ising \textit{anti}ferromagnet (AFM) in a \textit{uniform} external field on a two-dimensional (2D) square lattice. 
Using single-spin-flip Metropolis Monte Carlo dynamics, we perform instantaneous quenches from $T=\infty \to T_f$ (where $T_f\ll T_c$) and find that the field supports a pair of metastable magnetization plateaus whose lifetime diverges at low temperature. 
We describe the plateaus in terms of stable local spin configurations. 
Between the plateaus is a `valley of ergodicity' where the system eventually converges to the correct ground state. 

Monte Carlo (MC) works by drawing sample states from the Boltzmann distribution $e^{-\beta H}$. 
The starting point is typically some randomized (far from equilibrium) state and then updates are performed for some time without collecting data until the system has reached equilibrium. 
In practice, this process requires Monte Carlo updates that can move domain walls.
In our case, the field breaks these updates, producing local energy minima where no single spin can be flipped without increasing the energy. 
This is true even for very small fields which do not change the equilibrium ground state. 
Unlike the ferromagnet \cite{spirin2001t0,spirin2001,barros2009,olejarz2012,blanchard2013,mullick2017,godreche2018,denholm2019,denholm2020}, our system reaches a frozen state quickly, eliminating the need for long simulations. 
The AFM's frozen states are extremely numerous, such that the probability of reaching the true ground state vanishes rapidly (in contrast to the ferromagnet, where the ground state is reached most of the time \cite{spirin2001t0}). 
This system provides a simple case for understanding ergodicity breakdown in Monte Carlo more generally and may have useful parallels to other sticky problems such as the random-field Ising model \cite{belanger1991} and spin glasses \cite{binder1986}. 

\section{The Ising antiferromagnet}

The Ising antiferromagnet in an external field $h$ is defined: 
\begin{align}
H = J \sum \limits_{\braket{i,j}} \sigma_i \sigma_j - h \sum \limits_i \sigma_i,
\end{align}
where $\sigma_i=\pm 1$, $J=1$ and $\braket{i,j}$ represents a sum over nearest neighbors on an $L\times L$ (2D) square lattice with periodic boundary conditions. 
Hereafter we will set \mbox{$J=1$} and use dimensionless units. 
The equilibrium zero-temperature behavior\footnote{We refer to \textit{equilibrium} behavior to distinguish from out-of-equilibrium behavior after a quench or metastable states like the magnetization plateaus. } is quite simple: for $h=0$, there are two degenerate ground states composed of alternating up and down spins. 
The energy of each of these states is 
\begin{align}
E_g = -2JL^2. \label{eq:gs}
\end{align}
We define $m$ to be the average magnetization such that \mbox{$-1 \leq m \leq 1$}, 
\begin{align}
m \equiv \frac{1}{L^2} \sum \limits_i \sigma_i.
\end{align}
The field shifts the energy of a state with magnetization $m$ by $-h m L^2$.  
For \mbox{$|h|>h_s=4$}, the field is strong enough to drive a first-order phase transition to the fully polarized state. 
The magnetization is therefore \cite{muller1977}: 
\begin{align}
m(T=0,h) = 
\begin{cases}
-1		&	h<-4, \\
0		&	-4<h<4, \\
1		&	h>4.
\end{cases} \label{eq:eqm}
\end{align}
At finite temperature, there are thermal fluctuations that reduce the ordering and round off the step-function-like behavior of $m(h)$.
At sufficiently high temperatures \mbox{($T>T_c=2.27$)} the magnetic order is completely destroyed, even at $h=0$.  

The dynamics of the ferromagnetic case \cite{spirin2001,spirin2001t0,oliveira2006,grzegorz2009,barros2009,olejarz2012,blanchard2013,godreche2018,denholm2020,lopes2020} and variants \cite{mullick2017,denholm2019} have been well studied. 
Studies of quenches in the ferromagnet have established the existence of a limited set of stable striped states with straight domain walls \cite{spirin2001,spirin2001t0,oliveira2006,grzegorz2009,barros2009,olejarz2012}. 
In the absence of a field, the antiferromagnet can be mapped onto the ferromagnet by a simple transformation of flipping all spins on one sublattice. 
Adding a uniform external field breaks this transformation. 
There have been relatively few studies of the AFM Ising model in a uniform field: a handful of mostly theoretical papers \cite{sneddon1979,zittartz1980,zoltan1980,santos1985,blote1990} and some Monte Carlo studies \cite{binder1980,lourenco2016}, none of which have reported the plateaus that we will describe here. 

\section{Monte Carlo Dynamics}

Markov chain Monte Carlo (MCMC or simply Monte Carlo) is one of the most common numerical techniques in statistical physics (and beyond). 
At the core of MCMC is a \textit{Markov process}: a procedure for stochastically generating a sequence of states via a transition probability function $P(x\to x')$. 
In order to be a valid Markov process, $P(x\to x')$ must \textit{(i)} produce a \textit{stationary distribution} $\pi(x)$ such that the probability of occupying each state $x$ remains constant over time, 
\textit{(ii)} be aperiodic (nonrepeating), and \textit{(iii)} be ergodic (every state $x'$ can be reached from every other state $x$ in a finite number of steps \cite{newman1999}). 
Meeting all three of these conditions is required to ensure that the MC program correctly samples the Boltzmann (or other) distribution and that expectation values (e.g. $\braket{m}$) and their error bars accurately reflect the properties of the distribution without systematic error. 
Once a MC program has achieved this (typically after some transient) it is said to have reached \textit{equilibrium} and has no memory of its initial state. 

When designing a Monte Carlo program, condition \textit{(i)} can be met by enforcing the detailed balance condition, 
\begin{equation}
\pi(x)P(x\to x') = \pi(x') P(x'\to x),
\end{equation}
and condition \textit{(ii)} is satisfied by using random numbers. 
Condition \textit{(iii)} is more difficult to guarantee because ergodicity is not a trivial property of the transition probabilities (dynamics), but depends also on the parameters and---for practical purposes---on the amount of computer time available. 
In the context of Monte Carlo methods, ergodicity refers to the practical issue of whether all possible configurations are \textit{attainable} in a reasonable amount of time.\footnote{This is the common usage of ergodicity as it applies to Monte Carlo methods \cite[p. 159]{newman1999}\cite[p. 27]{landau2015}\cite{neirotti2000}.}
For example, single spin flip updates in the ferromagnet are ergodic, but below $T_c$ the time required to flip between the competing ground states rapidly diverges with size, and finite-time simulations will often remain stuck in either the $m>0$ or $m<0$ portion of the state space (common for symmetry-breaking transitions). 
More insidiously, simulations can become stuck in some non-obvious subspace where they exhibit apparently well-behaved dynamics and error bars, but nonetheless produce incorrect results. 

The difficulty is that Monte Carlo is used precisely when it is impossible to brute-force test all possible combinations $x,x'$ to ensure that they are connected by a finite number of steps.\footnote{For the Ising model, the number of states scales like $2^{L^2}$, so even for $L=100$ there are over over $10^{3010}$ states. For more complicated algorithms, such as quantum Monte Carlo, the state space can easily reach $10^{10^9}$ \cite[p. 112]{mythesis}.} 
In most cases, a well-designed Monte Carlo program running for a very long time will only visit a vanishingly small fraction of the full state space. 
Guaranteeing ergodicity is therefore impossible in most cases.\footnote{A trivial way to guarantee ergodicity would be to draw completely random spin configurations and do a weighted average using the Boltzmann weights. In this case, \mbox{$P(x\to x')=2^{(-L^2)}$} is constant. This procedure, however, would be extremely inefficient, since most random configurations have high energy and will make only exponentially small contributions to the average. }
Instead, one typically tests for ergodicity empirically by checking that the simulation appears to produce equilibrium behavior and that the autocorrelation time is short, but this does not \textit{guarantee} that it is sampling the full space. 

In this work, we use the standard Metropolis algorithm \cite{metropolis1953} with randomly-selected single spin-flip updates \cite{landau2015}. 
For each update, we select a spin $\sigma_i$ at random and flip it with probability 
\begin{align}
P = \min \left[ 1, e^{-\left(\sum \limits_j \sigma_j - h \right) \Delta \sigma_i /T} \right],\label{eq:metrop}
\end{align}
where $\sigma_j$ are its nearest neighbors. 
Updates that decrease the energy or leave it unchanged are always accepted and updates that increase the energy are accepted with \mbox{$P=e^{-\Delta E/ T}$} (similar, but not identical, to Glauber dynamics \cite{glauber1963}). 
Each Monte Carlo sweep (MCS) consists of $L^2$ of these attempted spin flips, and we will use the abbreviation kMCS to indicate units of $10^3$ MCS.  
We focus on instantaneous quenches from $T=\infty\to T_F$ by starting with a randomized initial state and performing MC updates at $T_F$. 
To facilitate replication of our work, we have made our complete Fortran program available online \cite{afmIsingCode}. 

Since the Ising model has no intrinsic physical dynamics, any Monte Carlo update scheme is necessarily artificial. 
Monte Carlo updates need not bear any resemblance to physical processes, since the goal is just to sample the state space according to the probability distribution. 
Therefore, \textit{simulation time} does not necessarily correspond to \textit{physical time} in any meaningful way.\footnote{\textit{Simulation time} is a measure of the number of Monte Carlo sweeps (MCS) that have been performed. } 
This is especially true for more complicated update schemes like cluster algorithms or loop updates in quantum Monte Carlo. 
Nonetheless, single spin flip updates \textit{do} resemble plausible physical dynamics and are often used in analogy to physical dynamics \cite{glauber1963,spirin2001t0,spirin2001,oliveira2006,hurtado2008,grzegorz2009,barros2009,olejarz2012,blanchard2013,mullick2017,godreche2018,denholm2020,lopes2020}. 
Similar physical dynamics could also be obtained by adding a small transverse field $h(\sigma^+ + \sigma^-)$ [although this would be the \textit{quantum} Ising model, a wholly different problem]. 
Furthermore, MCMC is a common technique and it is interesting to understand Monte Carlo dynamics in their own right. 
The breakdown of ergodicity we will describe here represents a simple way to understand non-ergodic behavior that occurs in more complicated Monte Carlo methods \cite{iaizzi2015,iaizzi2017,iaizzi2018metamag,mythesis} where the underlying mechanisms are more difficult to understand. 

\section{Observations}

\begin{figure} 
\includegraphics[width=83mm]{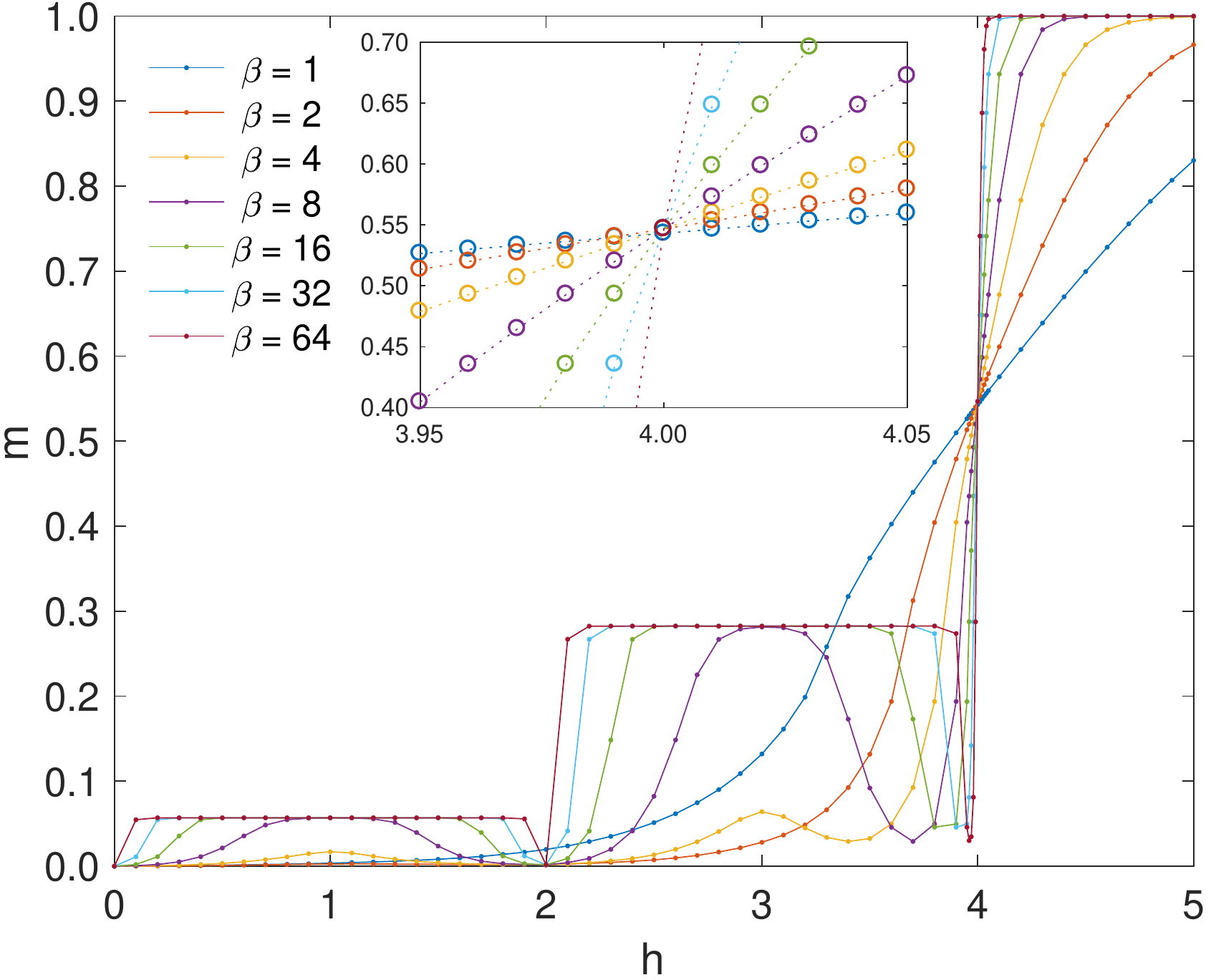}
\caption{Magnetization for instantaneous quenches from \mbox{$T=\infty$} followed by 40 kMCS at $\beta=1/T_{\rm F}$ for $L=512$. Each point is the average of 200 independent quenches; error bars are smaller than the markers. \underline{Inset:} Enlarged view around $h_s$ to show the vanishing $T$-dependence. \label{f:magcurve}}
\end{figure}

\begin{figure} 
\includegraphics[width=83mm]{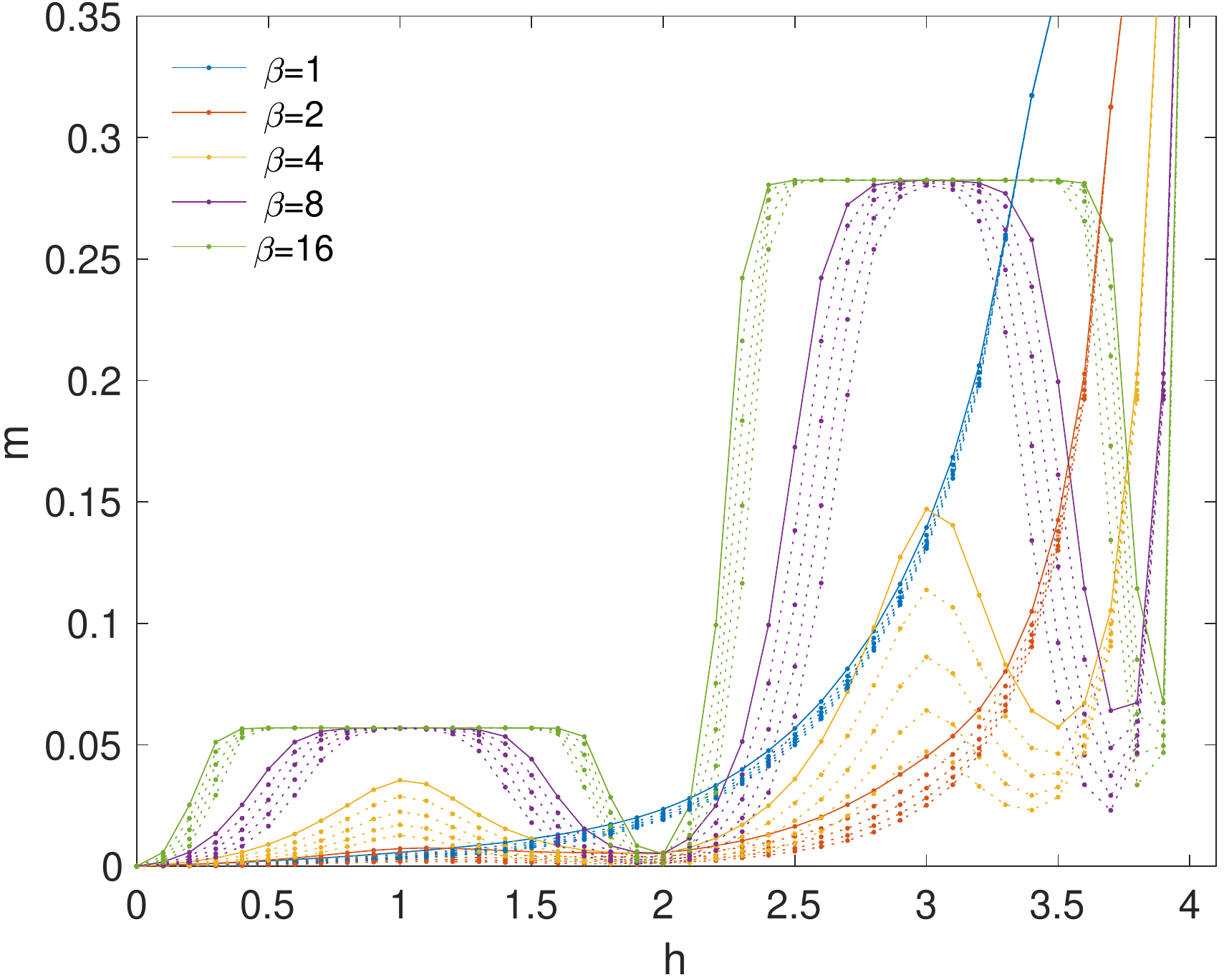}
\caption{Time dependence of magnetization for instantaneous quenches from \mbox{$T=\infty$} to $\beta=1/T_{\rm F}$ followed by 5 kMCS (solid lines), 10 kMCS, 20 kMCS, 40 kMCS, and 80 kMCS (same-color broken lines, descending) with $L=512$. Each point is the average of 200 independent quenches; error bars are smaller than the markers. \label{f:ftime}}
\end{figure}

In \cref{f:magcurve} we plot the magnetization $m$ resulting from instantaneous quenches to finite inverse temperature \mbox{$\beta=1/T$} at external field $h$ followed by 40 kMCS.\footnote{We will use the SI prefix k for brevity: 40 kMCS = $40 \times 10^3$ Monte Carlo sweeps. } 
At the highest temperature here $(T=1<T_c)$ the magnetization behaves as expected for equilibrium, with the finite temperature rounding off the sharp edges in the zero-temperature curve [\cref{eq:eqm}]. 
At lower temperatures the behavior is quite unusual. 
The magnetization develops two plateaus that become progressively sharper as $T\to 0$. 
Unlike conventional magnetization plateaus, these do not pass through the equilibrium zero-temperature magnetization curve, but are instead substantially higher. 
For \mbox{$0<h<2$} there is a plateau at $m_1 \approx 0.0569$ and for \mbox{$2<h<4$} there is a plateau at $m_2 \approx 0.283$. 
These frozen states do not break any obvious symmetries of the system (see \cref{f:h1000config,f:h3000config}); the number of these states grows rapidly (probably exponentially) in $L$.

The first signs of the plateaus appear around \mbox{$\beta=4$} and they are well-defined by \mbox{$\beta=16$}. 
Stranger still, the correct ground state is \textit{restored} in `valleys of ergodicity' between the plateaus at $h=0,2$. 
These valleys become narrower as \mbox{$T\to 0$}. 
Near and slightly below the $h_s$, ergodicity is at least partially restored; this third valley of ergodicity becomes narrower and closer to $h_s$ as $T \rightarrow 0$. 
At $h_s$, the $T$-dependence vanished rapidly (as seen in the inset), which will be discussed further in \cref{s:hsat}. 

The magnetization plateaus in \cref{f:magcurve} are not merely the result of \textit{slow} equilibration, but completely frozen dynamics. 
We explore the time dependence further in \cref{f:ftime}. 
For each $\beta$, we show the results of a quench followed by 5 kMCS, 10 kMCS, 20 kMCS, 40 kMCS and 80 kMCS. 
The central, flat regions of the plateaus constitute the \textit{strongly-frozen} regime, where 
a frozen state is reached quickly and there is no further progress.\footnote{Often very few ($<10$) MCS are required to reach a strongly-frozen state.} 
As we lower the temperature, the plateaus grow from their centers $(h=1,3)$ and become wider as $T\to 0$. 
The temperature controls the width of the flat strongly-frozen region of the plateaus, but once in a frozen state, the temperature itself is effectively irrelevant. 

Outside of the strongly-frozen regime are the valleys of ergodicity, centered around the ergodic points $(h=0,2)$. 
At the ergodic points, the relaxation to the correct ground state is fast, but as we move away from these points the relaxation time grows rapidly, becoming effectively infinite in in the strongly-frozen regime. 
This slow relaxation time explains the slopes of the valleys of ergodicity and appears as a clear time dependence in those regions in \cref{f:ftime} (whereas the ergodic points and strongly-frozen regimes are fully converged by 5 kMCS). 
The nature of this slow relaxation and the exact boundary between this regime and the strongly-frozen regime are interesting, but we do not address them further in the present study. 

The valleys of ergodicity become progressively narrower as $T\to0$, and for $T=0$ we expect that the correct ground state will only be reached at exactly $h=0,2$. 
Extrapolating from \cref{f:magcurve}, we can predict the form of the magnetization for quenches to exactly zero temperature:
\begin{align}
m(T=0,h) = 
\begin{cases}
0		&	h=0, \\
0.0569	&	0<h<2J, \\
0		&	h = 2J, \\
0.283	&	2J<h<4J, \\
0.5467	&	h=h_s=4J, \\
1		&	h>h_s,
\end{cases}
\end{align}
which is dramatically different from the equilibrium behavior [\cref{eq:eqm}]. 
(The actual values used here are from \cref{tab:pop}). 
 
Hereafter we will focus on instantaneous quenches from $T=\infty$ to the strongly-frozen regime. 
In this regime, finite-size effects vanish quickly. 
Heuristically, this is the result of the coarsening process \cite{bray1994,cugliandolo2015} halting  while the correlation length is still short. 
Finite-size effects are discussed in \cref{s:fss}. 
We will also restrict our analysis to systems with periodic boundary conditions (PBC). 
The case of open boundary conditions (OBC), which is largely the same except for the presence of prominent finite-size effects, is described in \cref{s:obc}. 

\section{Local configurations \label{s:conf}}

\begin{figure} 
\includegraphics[width=55mm]{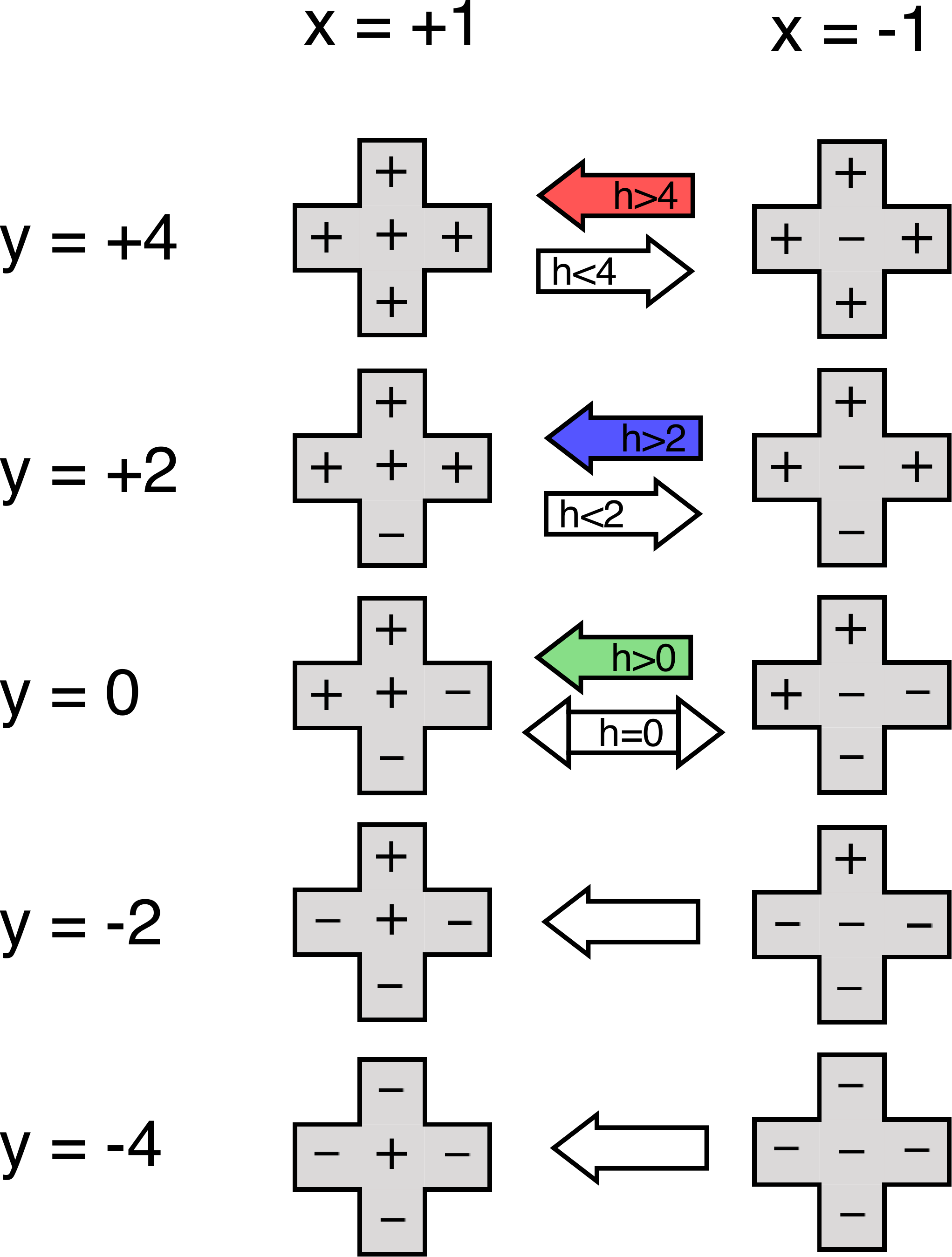}
\caption{Schematic of all possible local spin configurations $C^y_x$ [\cref{eq:cdef}] with center spin \mbox{$x=\sigma_i=\pm1$} and nearest neighbors \mbox{$y=\sum \sigma_j = 0,\pm 2 \pm4$} [\cref{eq:xy}]. In the updates, the neighbors are treated as fixed and the center spin is flipped \mbox{$C^y_x\to C^y_{-x}$}. For each pair (row) $C^y_{\pm 1}$, the configuration with lower energy is stable and the other is unstable (at \mbox{$T=0$}). If they are degenerate, the update is always accepted. For $h=0$, the transitions are described by the white arrows; the stable state is for $x$ to be antiparallel to $y$, and when $y=0$, $C^0_{\pm1}$ are degenerate. For $h>0$, $C^0_{-1}$ becomes unstable and always flips to $C^0_{+1}$ (as indicated by the green arrow). As $h$ increases, it is able to overcome the effects of the neighbors. For $h>2$, the direction of the $C^{+2}_{\pm1}$ transition changes (as described by the blue arrow). Above $h=4$, the spin always flips to $(+)$ and only $C^y_{+1}$ is stable (red arrow). \label{f:config}}
\end{figure}

The magnetization plateaus do not reflect any ordering or symmetry-breaking transition. 
Instead, they are an out-of-equilibrium phenomenon that can be understood in terms of local spin configurations that are stable under our dynamics. 
We will set $J=1$ and define $x,y$: 
\begin{subequations}
\begin{align} 
x = & \sigma_i = \pm 1 \\
y = & \sum \limits_j \sigma_j = 0, \pm 2, \pm 4,
\end{align}\label{eq:xy}%
\end{subequations}%
where $\sigma_i$ is the spin that we will attempt to flip and $\{ \sigma_j \}$ are its nearest neighbors (which should be considered fixed). 
We will describe these local spin states using the notation:
\begin{align}
C^y_x. \label{eq:cdef}
\end{align}
In \cref{f:config} we show 10 local spin configurations that describe all possible combinations of \mbox{$x=\pm1$} and \mbox{$y=0,\pm2,\pm4$} (other configurations are reachable by rotations and permutations of the neighbors). 

In the language of these local states, the Metropolis algorithm chooses a random spin, which is at the center of a configuration $C^y_x$ and attempts to flip it to $C^y_{-x}$. 
This results in a change of energy 
\begin{align}
\Delta E = (y-h)\Delta x = -2(y-h)x.
\end{align}
From \cref{eq:metrop}, the probability of accepting this spin flip is
\begin{align}
P =  \min \left[1, e^{-\Delta E/T} \right] = \min \left[1, e^{-(y-h)\Delta x/T} \right].\label{eq:pflip}
\end{align}
At zero temperature the updates are even simpler: changes are accepted if and only if \mbox{$ E(C^y_{-x})\leq E(C^y_{x})$}. 

When $y=h$, $C^y_x$ and $C^y_{-x}$ are degenerate, so $\Delta E =0$. 
\mbox{$\Delta E=0$} updates are special because they are \textit{reversible} (they can be undone), whereas reversing a $\Delta E<0$ update requires a $\Delta E >0$ update, which is impossible at $T=0$.\footnote{These reversible ($\Delta E =0$) updates are also called `active spins' by Ref.~\onlinecite{barros2009}, and `flippable states' by Ref.~\onlinecite{olejarz2012}. } 
Reversible updates are only present when $h$ is tuned to one of the five possible values of \mbox{$y=0,\pm2,\pm4$}. 
These values of $h$ correspond to the valleys of ergodicity observed in \cref{f:magcurve}. 
For all $y \neq h$, each pair $C^y_{+1}$ and $C^y_{-1}$ has one stable and one unstable state. 

At finite temperature, all updates are (strictly) reversible, but updates that increase the energy are exponentially suppressed. 
At sufficiently low temperatures the suppression becomes so strong that updates are irreversible on any practical timescale. 
For example, when $h=1$, $\beta=16$, the probability of flipping a stable $C^0_{+1}$ to higher-energy $C^0_{-1}$ is $e^{-2\times 16}\approx 10^{-14}$ [see \cref{eq:pflip}]. 
Therefore, approximately $10^{11}$ MCS would have to be performed before one of these updates it likely to be accepted (for a $100\times 100$ system), which is over one million times the length of the longest simulations considered here. 
In plateau states, every site occupies one of these stable local configuration from which updates are exponentially suppressed. 
In the strongly-frozen regime (the central flat part of the plateau), the numerical results demonstrate that this probability is low enough to stabilize the plateaus for extremely long times. 
Even very long simulations were never observed to escape from a plateau state in the strongly-frozen regime. 

\section{Explanation of the Plateaus \label{s:plateau}}

We can describe spin states in terms of local spin configurations (\cref{f:config}). 
Configurations $C^{\pm 4}_{\mp 1}$ correspond to bulk AFM ground states, which have one sublattice fully occupied by $C^{+4}_{-1}$ and the other occupied by $C^{-4}_{+1}$. 
$C^{\pm2}_{\mp1}$ correspond to horizontal and vertical domain walls [see \cref{f:h1000config}]. 
In frozen states, every single spin is at the center of one of several stable local configurations. 
Except where explicitly stated otherwise, this section will describe the zero temperature limit. 

\subsection{$T=\infty$}

Our quenches start from a randomized initial spin state corresponding to $T=\infty$. 
All $C^{0,\pm2,\pm4}_{\pm 1}$ are stable. 
The expected proportions can be derived from simple combinatorics (see \cref{s:tinf}). 
The numerical results in \cref{tab:pop} validate these predictions.

\subsection{$h=0$}

At $h=0$, the system can be mapped exactly onto the Ising ferromagnet \cite{spirin2001}. 
The stable local configurations are $C^{+4}_{-1}$, $C^{+2}_{-1}$, $C^{-2}_{+1}$ and $C^{-4}_{+1}$, while $C^0_{+1}$ and $C^0_{-1}$ are degenerate and switching between them is reversible. 
In terms of domains, the stability of $C^{+4}_{-1}$ and $C^{-4}_{+1}$ means bulk AFM domains are stable, and the stability of $ C^{+2}_{-1}$ and $C^{-2}_{+1}$ makes straight-line domain walls stable as well. 
Domain wall corners $(C^0_{\pm1})$ are unstable. 
This means that even at zero temperature, there are reversible updates that move domain walls and make it possible to reach the ground state in finite time. 
In practice, after an instantaneous quench to $T=0$ the system will become stuck in a stable stripe state \cite{spirin2001,spirin2001t0,barros2009,olejarz2012} with probability $P=0.3390...$, which can be derived from a connection to continuum percolation at the critical point \cite{barros2009}.  
The domain walls in these stripe states account for the \mbox{$\approx 2\%$} of $C^{\pm2}_{\mp1}$ states in \cref{tab:pop}.\footnote{The final state has a stripe with probability $P\approx0.3390$ \cite{barros2009}. With periodic boundary conditions, a single stripe has two domain walls of length $L$, each with $2L$ $C^{\pm2}_{\pm1}$ states, so we expect $(0.3390)\times2\times2L/L^2\approx 2\%$ of local states to be $C^{\pm2}_{\pm1}$ for $L=64$.} 
Note that the AFM ground states and the stripe defect states are all  frozen states under these dynamics. 

\subsection{First plateau}

\begin{figure} 
\includegraphics[width=60mm]{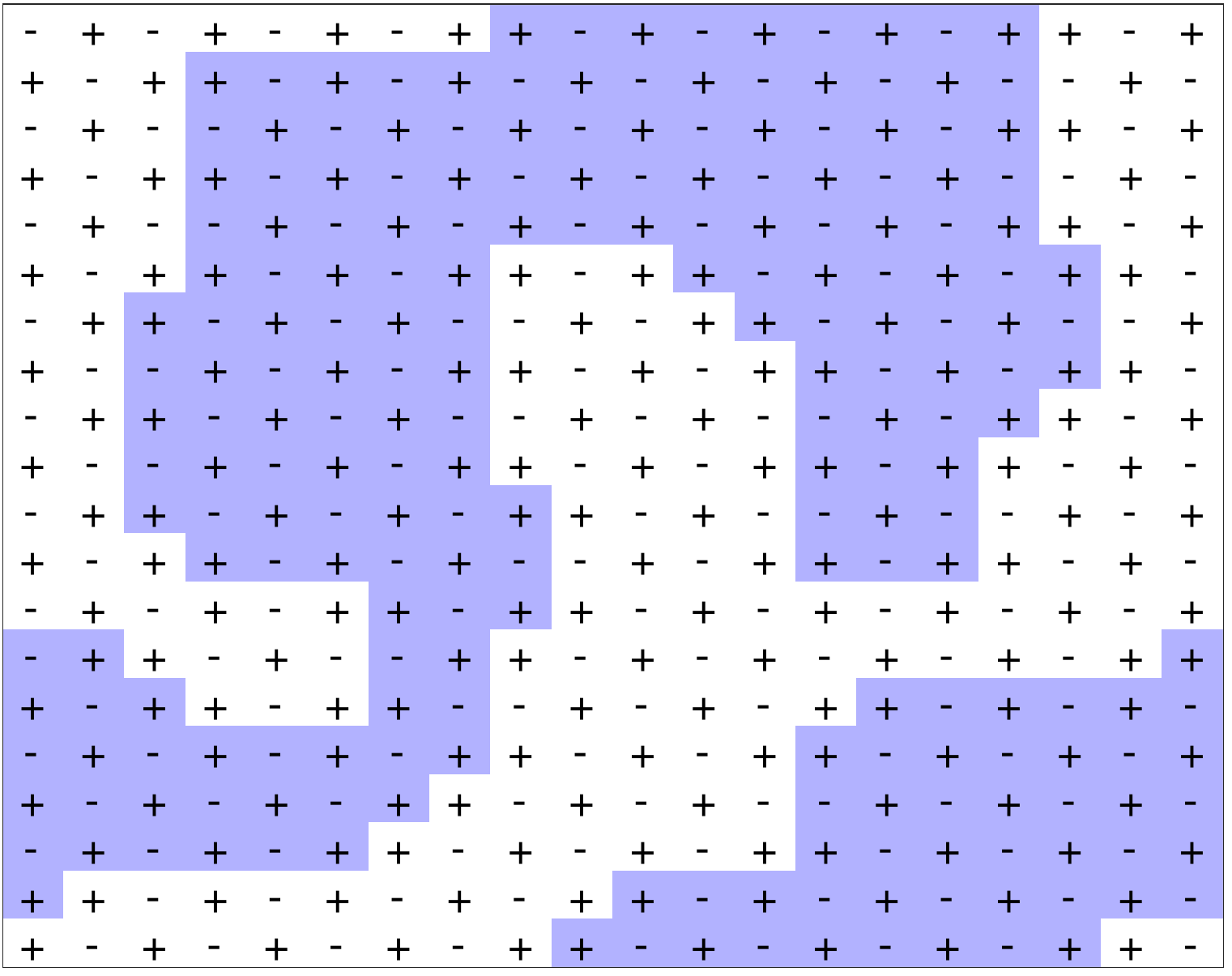}
\caption{Example $m_1$ plateau state generated by an instantaneous quench from \mbox{$T=\infty \to 0$} with $h=1$ and $L=20$.  Markers $\pm$ indicate spin, and the background shading shows the AFM domains. Each spin is at the center of a stable local configuration and no further updates are possible. Excess $(+)$ spin is located at domain wall corners. \label{f:h1000config}}
\end{figure}

The first $(m_1)$ plateau occurs for $0<h<2$. 
The field breaks the degeneracy between $C^0_{+1}$ and $C^0_{-1}$, so now only $C^0_{+1}$ is stable and there are no reversible local spin flips. 
Bulk domains and straight domain walls remain stable, but now corners and diagonal domain walls with excess $(+)$~spin are stable as well, giving rise to a net magnetization. 
The $m_1$ plateau is composed of an ensemble of all states that obey these domain wall rules.
In \cref{f:h1000config}, we show an example of an $m_1$ plateau state. 

The initial random state is a mix of all $C^y_x$. 
As the simulation progresses, it eventually flips all the unstable states ($C^{+4}_{+1}, C^{+2}_{+1}, C^{0}_{-1}, C^{-2}_{-1}, C^{-4}_{-1}$), leaving only stable local states ($C^{+4}_{-1}, C^{+2}_{-1}, C^{0}_{+1}, C^{-2}_{+1}, C^{-4}_{+1}$). 
In \cref{tab:pop} we show the result of averaging over many realizations of these frozen plateau states; only the expected stable local configurations are present. 
Once the system is composed of only stable local spin configurations, no further updates are possible since any single spin flip would raise the energy and be rejected. 
Two of these stable states are the true AFM ground states, but almost all initial states will intersect with some other stable state first and become permanently stuck there.  
There are so many plateau states that the ground state is never reached in practice. 

In \cref{f:maghist}(a), we show a histogram of the magnetization in the $m_1$ plateau. 
The distribution of magnetizations within the frozen plateau states is very narrow and does not overlap with the ground state or the other plateau.  
The excess energy above the ground state [\cref{f:enrghist}(a)] is also narrowly distributed about a mean. 

\begin{figure} 
\includegraphics[width=60mm]{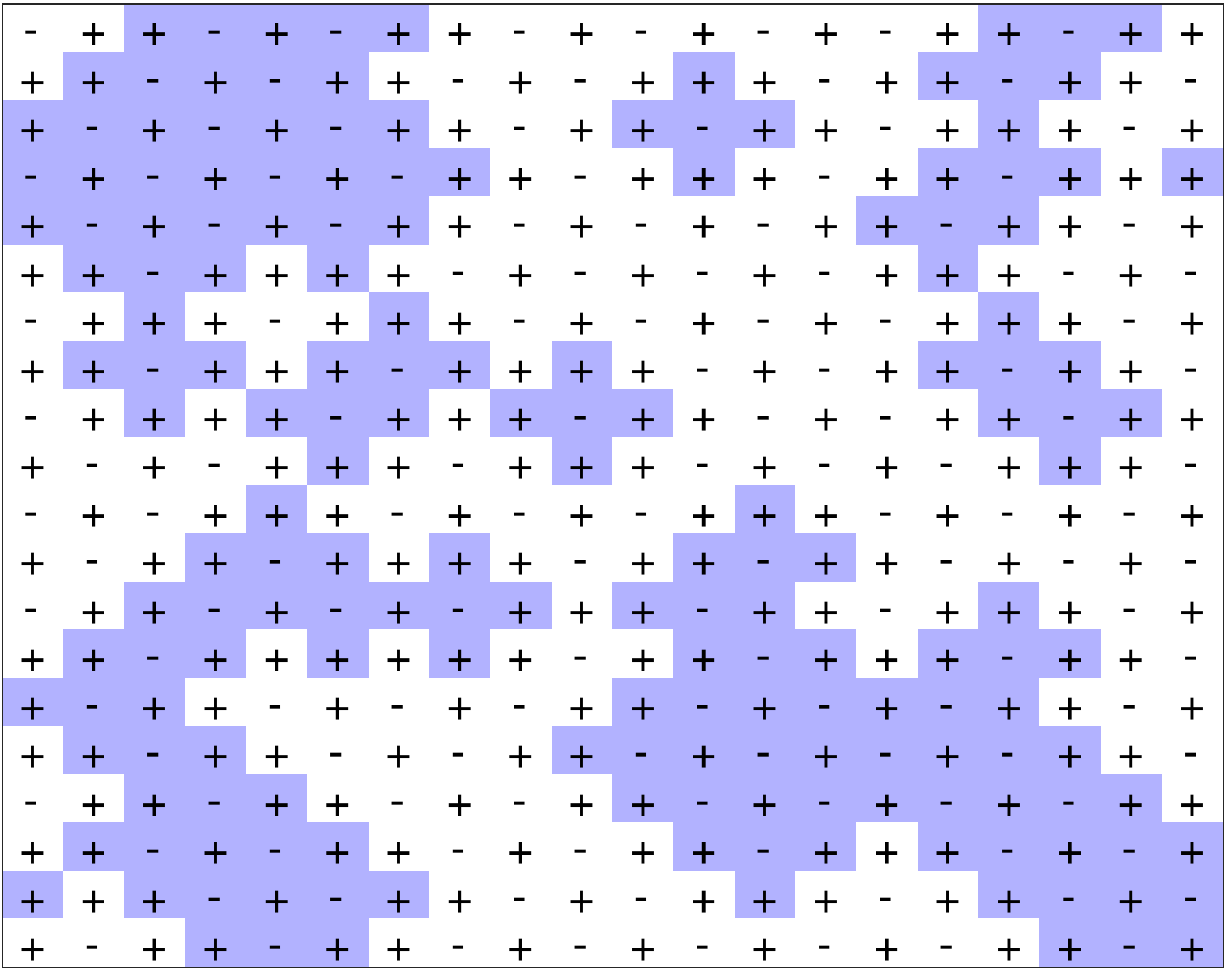}
\caption{Example $m_2$ plateau state generated by an instantaneous quench from \mbox{$T=\infty \to 0$} with $h=3$ and $L=20$. Markers $\pm$ indicate spin, and the background shading shows the AFM domains. Each spin is at the center of a stable local configuration and no further updates are possible. Excess $(+)$ spin is located along the domain walls.  \label{f:h3000config}}
\end{figure}

\subsection{$h=2$}

Around $h=2$, the simulation recovers ergodicity. 
There is now a $\Delta E=0$ update: $C^{+2}_{+1} \leftrightarrow C^{+2}_{-1}$. 
The presence of this reversible update allows free movement of domain walls and makes it possible to reach the true ground state even for zero temperature quenches (although the time required to do so can be very long). 
For $T=0$, ergodicity is only recovered at exactly $h=2$, but for finite temperature there is a valley of ergodicity centered around $h=2$ which becomes broader at higher temperatures. 

Quenches with $h=2$ are actually slightly better at finding the true AFM ground state than quenches with $h=0$ because the stripe defects \cite{spirin2001} that appear at $h=0$ are no longer stable. 
It \textit{is} possible to become stuck in an analogous diagonal stripe state, but this is much less common. 
These diagonal stripe defects appear in \cref{tab:pop} as a small population of $C^0_{+1}$. 
The stability of these diagonal stripes is likely affected by the aspect ratio \cite{barros2009}; it is therefore possible that for some non-square system one could guarantee reaching the ground state for $h=2$, although we have not investigated that here. 

\subsection{Second plateau}

\begin{figure} 
\includegraphics[width=80mm]{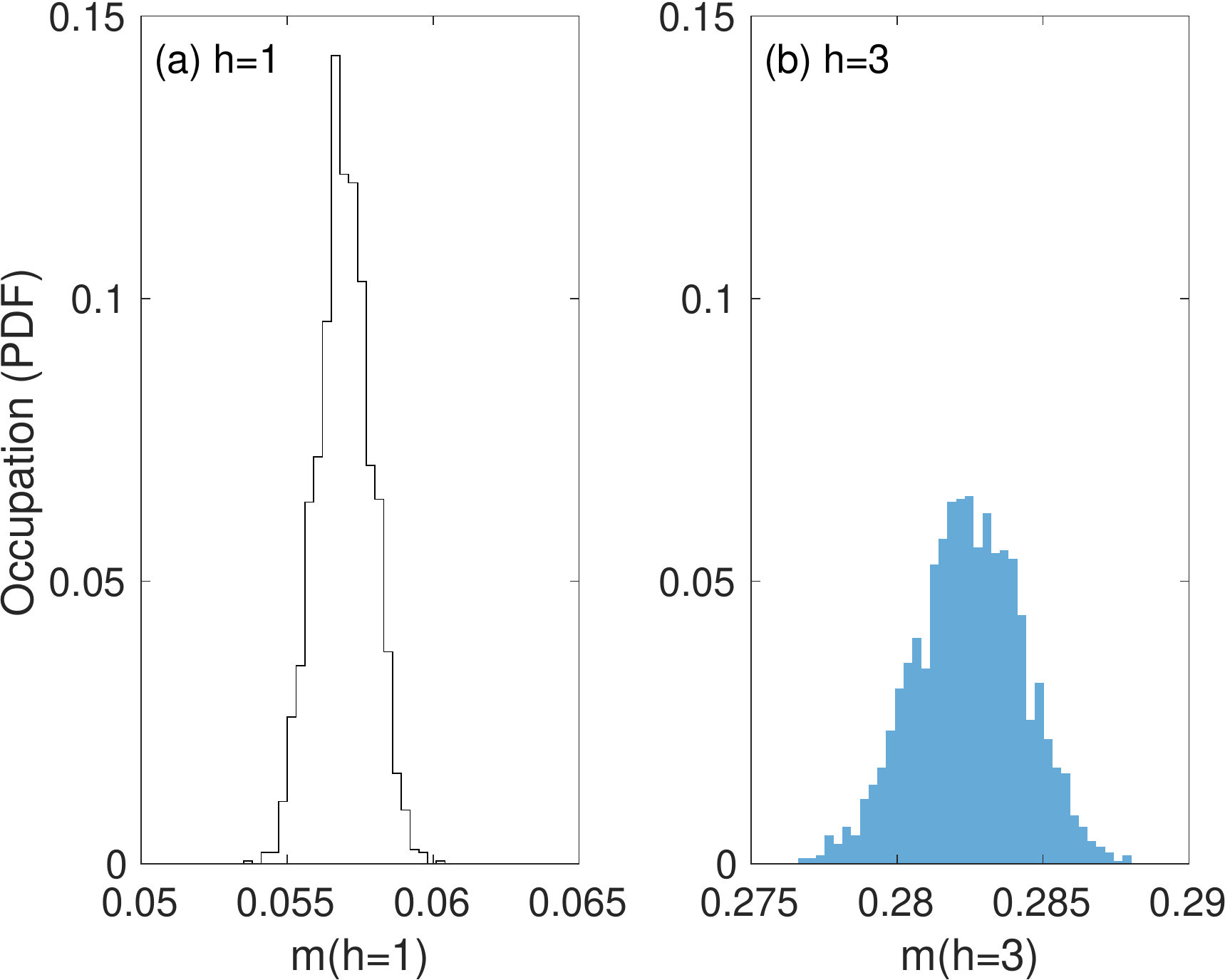}
\caption{Magnetization histograms for the \textbf{(a)} $m_1$ and \textbf{(b)} $m_2$ plateaus resulting from 2,000 independent instantaneous quenches from $T=\infty \to 0$ for a $256\times256$ system with $h=1$ and $h=3$, respectively. Measurements were taken after reaching $P_{\rm accept}=0$ for all sites [\cref{eq:pflip}]. Both panels use the same bin width. These distributions are narrow and well-separated both from each other and from the ground state. \label{f:maghist}}
\end{figure}

\begin{figure} 
\includegraphics[width=80mm]{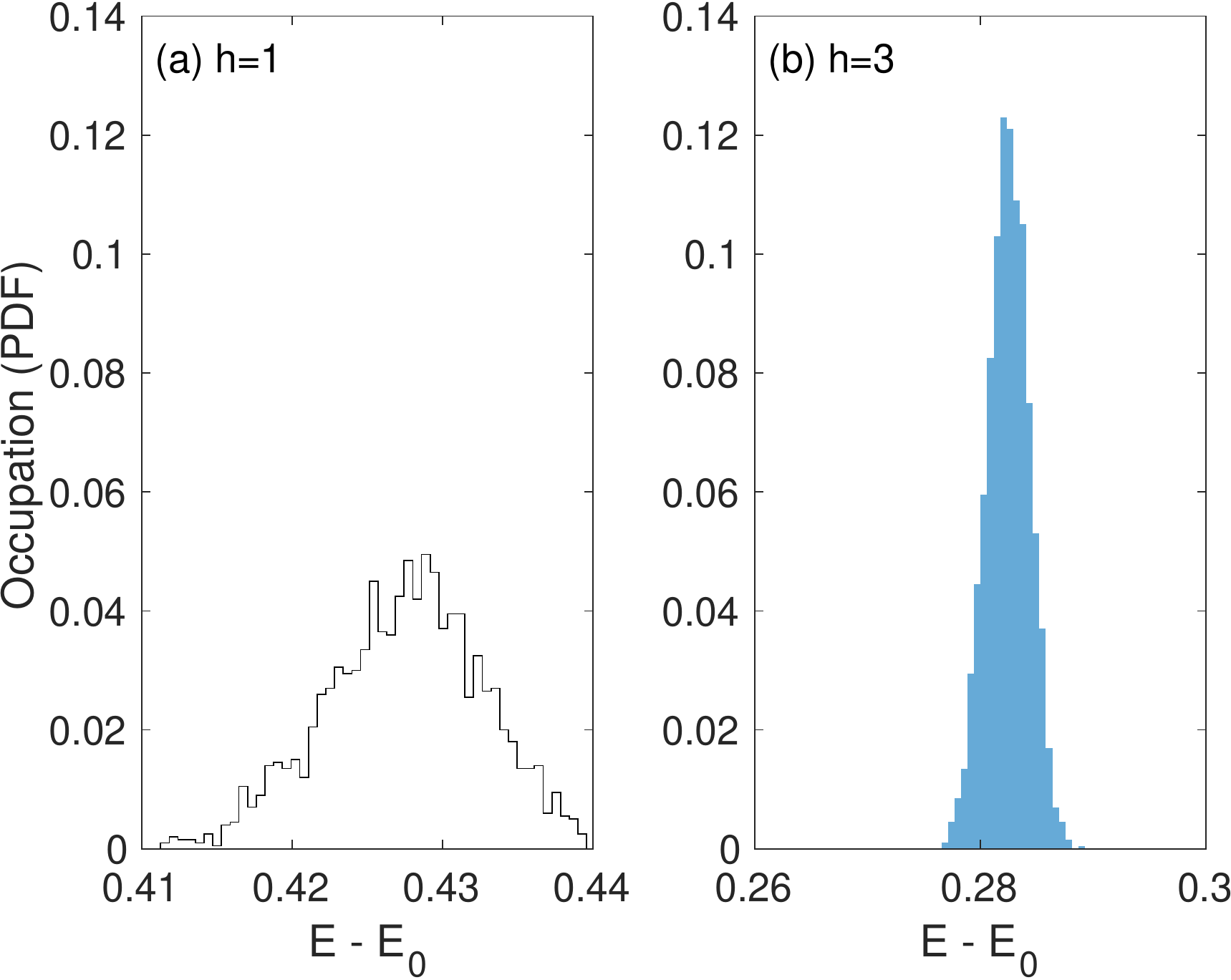}
\caption{Distribution of excess energy per site [relative to the ground state, \cref{eq:gs}] in  the \textbf{(a)} $m_1$ and \textbf{(b)} $m_2$ plateaus resulting from 2,000 independent instantaneous quenches from $T=\infty \to 0$ for a $256\times256$ system with $h=1$ and $h=3$, respectively (see also \cref{f:maghist}). Simulations were run until the acceptance probability was zero for all sites. The bins are the same width in both panels. Note that the excess energy for $h=3$ is \textit{lower} than for $h=1$ even though the $h=3$ plateau is further from equilibrium. \label{f:enrghist}}
\end{figure}

In the second $(m_2)$ plateau \mbox{$(2<h<4)$} the stable local states are $C^{+4}_{-1}, C^{+2}_{+1}, C^{0}_{+1}, C^{-2}_{+1}, C^{-4}_{+1}$ (see also \cref{tab:pop}). 
An example of such a configuration can be seen in \cref{f:h3000config}. 
In the $m_2$ plateau, bulk domains are still stable, but straight domain walls are not. 
Only diagonal domain walls are stable, and these host the excess $(+)$ spin, causing the net magnetization of $m_2\approx 0.283$. 

Similar to the $m_1$ plateau, the magnetization [\cref{f:maghist}(b)] and energy [\cref{f:enrghist}(b)] in the plateau states are narrowly-distributed and well-separated from both the ground state and the other plateau. 
Somewhat counterintuitively, the energy of the $m_2$ plateau is actually lower than the $m_1$ plateau, even though it is further from the correct ground state.  

\subsection{$h=h_s$ \label{s:hsat}}

\begin{figure} 
\includegraphics[width=60mm]{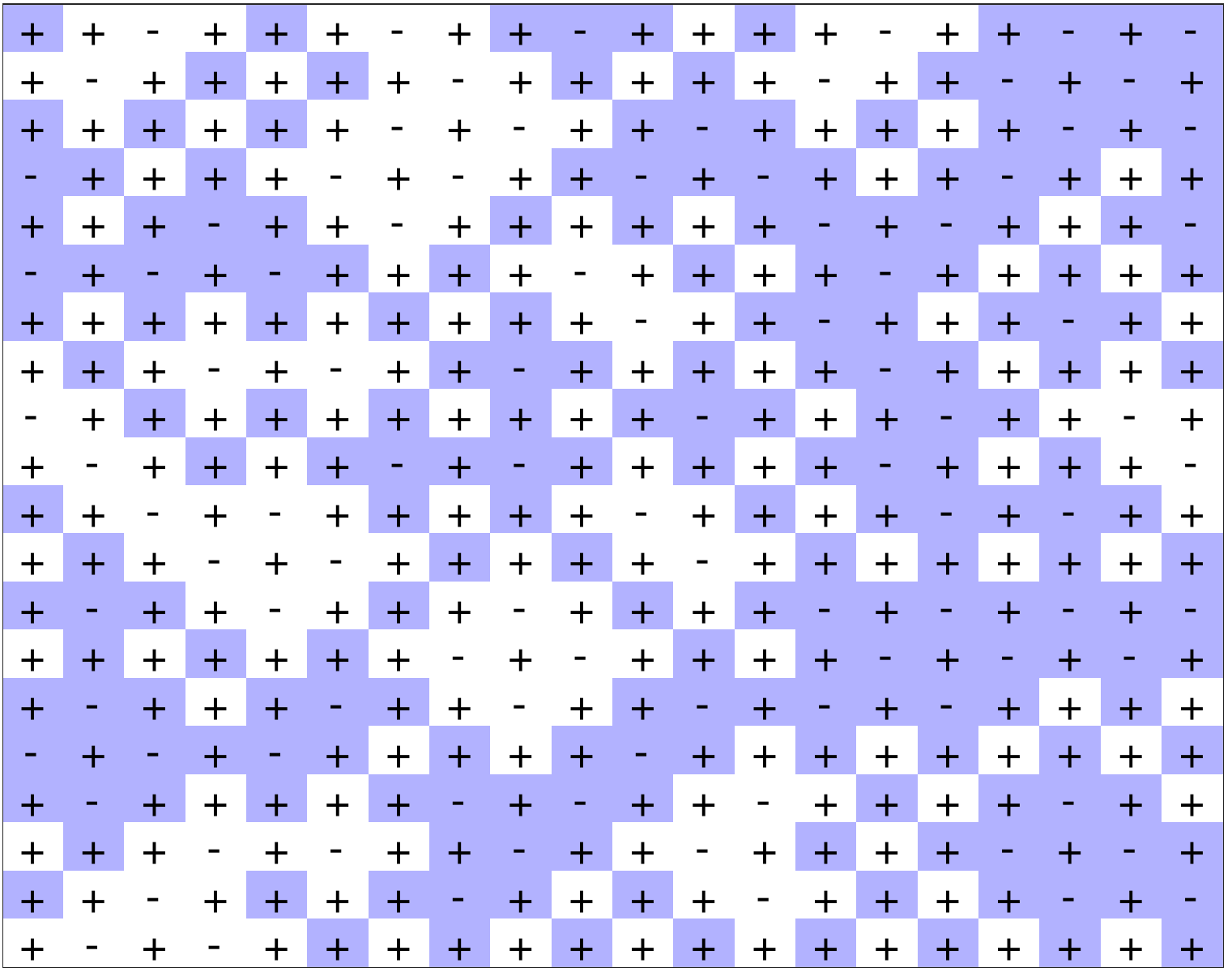}
\caption{Example configuration for the saturation point generated by an instantaneous quench from \mbox{$T=\infty \to 0$} with \mbox{$h=h_s=4$} and $L=20$. Markers $\pm$ indicate spin, and the background shading shows the AFM domains. Note: this is not a frozen state. \label{f:h4000config}}
\end{figure}

At $h=h_s$, there is no freezing, but the behavior is still unusual. 
Notably, the temperature dependence vanishes rapidly as $T\to 0$ with \mbox{$m(h=h_s)\to 0.5467$}. 
Even at zero temperature, the simulation does not freeze; it instead samples a highly-degenerate manifold of ground states, where there is coexistence of the fully polarized state and both AFM ground states. 
We show an example of such a state in \cref{f:h4000config}: there are patches of both AFM ground states as well as fully-polarized areas. 
For $T=0$, the simulation samples a range of magnetizations, but the energy always converges to the exact ground state [\cref{eq:gs}]. 
The behavior at $h=h_s$ can be mapped onto a `reversible random sequential adsorption process' \cite[p.~220]{krapivsky2010} or the zero-mobility hard squares problem \cite{huckaby1984,fernandes2007,liu2000,hu1989}, which are described in \cref{s:hardsquares}. 

\begin{figure} 
\includegraphics[width=80mm]{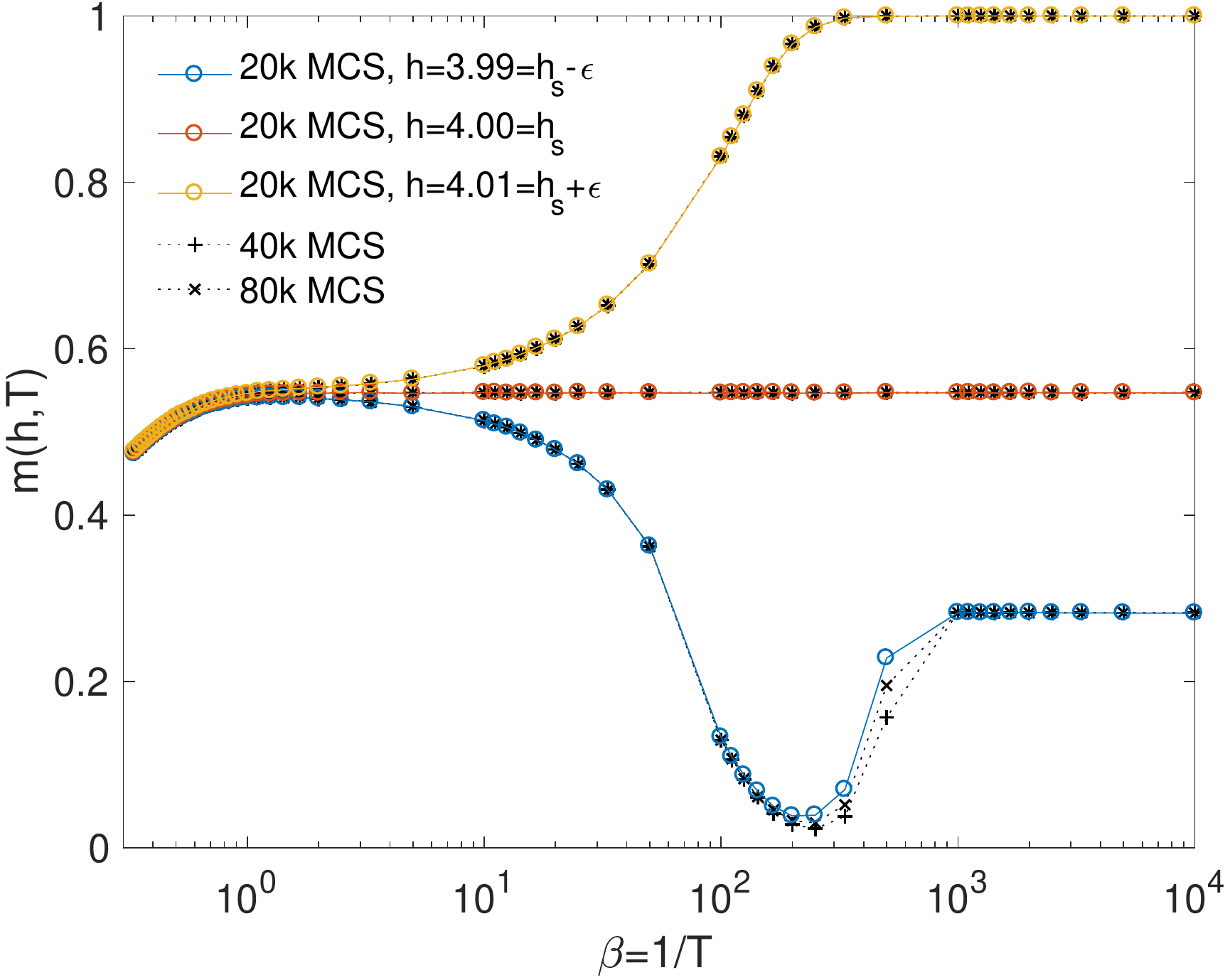}
\caption{Magnetization of a $128\times 128$ system with $h=h_s$ compared to \mbox{$h=h_s\pm \epsilon$} after instantaneous quenches from \mbox{$T_0=\infty$} to \mbox{$\beta=1/T$} followed by 20 kMCS ($\circ$), 40 kMCS ($\times$) and 80 kMCS ($+$). Each point is an average of 200 independent quenches. Error bars are smaller than the markers. At $h=h_s$ the finite temperature effects vanish rapidly and \mbox{$m(h_{s}) = 0.5469$} for all \mbox{$T<10^{-3}$}. \label{f:hsat}}
\end{figure}

In \cref{f:hsat}, we compare the temperature scaling of magnetization for quenches at $h_s$ to quenches at $h_s\pm \epsilon$ (where \mbox{$\epsilon = 0.01$}).  
At $h_{s}$, the $T$-dependence vanishes rapidly. 
Small deviations from $h_s$, however, cause large temperature effects. 
For \mbox{$h_s+\epsilon$}, the magnetization quickly converges to saturation \mbox{$(m=1)$}.  
For \mbox{$h_s-\epsilon$}, the behavior is more interesting---the magnetization first decreases (corresponding to the valley of ergodicity just below $h_s$ [\cref{f:magcurve}]) and then increases to $m_2$ as the temperature becomes low enough for \mbox{$h_s-\epsilon$} to lie in the $m_2$ plateau. 
Figure \ref{f:hsat} includes the magnetization at 20 kMCS, 40 kMCS and 80 kMCS after the quench. 
In most cases, the time evolution has finished before 20 kMCS, and the three lines coincide. 
The only deviation occurs for $h=3.99$ in a small window $10^2 < \beta < 10^3$. 
In this range the system is on the edge of the $m_2$ plateau and there is a longer relaxation time (see \cref{f:ftime}).   

The reason for the vanishing $T$-dependence has nothing to do with the conventional Ising ordering transition (which occurs at \mbox{$T_c\approx2.27$}); it is a property of our dynamics. 
Each $(-)$ spin is surrounded by four $(+)$ spins $(C^{+4}_{-1})$, but $(+)$ spins can have any number of parallel neighbors, so there are six stable local spin configurations: the degenerate pair $C^{+4}_{\pm1}$ and four other states with $x=+1$, $y\neq4$: $C^{+2}_{+1}, C^{+4}_{0}, C^{+4}_{+1}, C^{-2}_{+1}, C^{-4}_{+1}$ [see \cref{tab:pop}]. 
The $C^{y\neq4}_{+1}$ states make up $\approx$55\% of the local configurations, but the likelihood of flipping them is exponentially suppressed in $1/T$,
\begin{align}
P(C^y_{+1}\to C^y_{-1}) =  e^{2(y-4)/T}= e^{-8/T}e^{2y/T},
\end{align} 
so their contribution to the dynamics is negligible. 
For example, consider $C^{+2}_{+1}$: at $T=1$ the update acceptance probability for this state is already very low \mbox{$(P=0.018)$}. 
At 28\% of the population, only about 1\% of accepted flips at $T=1$ will be $C^{+2}_{+1}\to C^{+2}_{-1}$, so its contribution is weak and decreasing as $e^{-4/T}$. 
Flipping $C^{0,-2,-4}_{+1}$ are suppressed even more strongly. 
The dominant contribution to the dynamics comes from flipping between the degenerate $C^{+4}_{+1}$ and $C^{+4}_{-1}$ states, which are always-accepted $\Delta E = 0$ updates (that do not depend on $T$). 
Combined with the exponential suppression of all other updates, the result is a vanishing $T$-dependence. 

\begin{table}[]
\begin{tabular}{l|r|r|r|r|r|r}
Config. 		& $h=0$ & $0<h<2$ & $h=2$ & $2<h<4$ & $h=4$ & $T=\infty$ \\ \hline
\# Samp.		& 2,000	& 2,000	  & 8,000 & 2,000	& 2,000 & 7,000 \\ \hline\hline
$C^{-4}_{-1}$ 	& 0 	& 0 	& 0 	& 0 	& 0 	& 3.12 \\ \hline
$C^{-4}_{+1}$ 	& 49.0 	& 28.7	& 49.8 	& 11.1	& 1.3 	& 3.13 \\ \hline
$C^{-2}_{-1}$ 	& 0 	& 0 	& 0		& 0 	& 0 	& 12.50	\\ \hline
$C^{-2}_{+1}$ 	& 1.0 	& 12.7	& 0		& 11.3 	& 6.9 	& 12.50	\\ \hline
$C^{0}_{-1}$  	& 0 	& 0 	& 0 	& 0 	& 0 	& 18.74	\\ \hline
$C^{0}_{+1}$  	& 0 	& 11.5 	& 0.2 	& 23.7 	& 18.3 	& 18.74	\\ \hline
$C^{+2}_{-1}$ 	& 1.0 	& 12.9 	& 0 	& 0 	& 0 	& 12.51	\\ \hline
$C^{+2}_{+1}$ 	& 0 	& 0 	& 0 	& 18.1 	& 28.3 	& 12.51	\\ \hline
$C^{+4}_{-1}$ 	& 49.0 	& 34.3 	& 49.9 	& 35.9 	& 22.7 	& 3.12	\\ \hline 
$C^{+4}_{+1}$ 	& 0 	& 0 	& 0 	& 0 	& 22.6  & 3.13	\\ \hline \hline
$\braket{m}$  	& 0 	& 0.0569& 0.0012& 0.283 & 0.5467 & 0.000	
\end{tabular}
\caption{Populations (\%) of local spin configurations [\cref{f:config}, \cref{eq:cdef}] for \mbox{$64\times64$} systems after quenches to $T=0$ with $h=0,1,2,3,4$ (averaged over many independent final states). For $h\neq4$, simulations were run until $P_{\rm accept} = 0$ for all sites [\cref{eq:pflip}]; for $h=4$, 50 kMCS were performed before taking a measurement. The rightmost column is sampled from (randomized) $T=\infty$ states. The final row is average magnetization. All quantities have statistical error of less than one unit in the last digit. \label{tab:pop}}
\end{table}

\section{Conclusions \label{s:discussion}}

We have examined quenches to low temperature in the 2D square-lattice Ising antiferromagnet using single-spin-flip Metropolis algorithm dynamics and showed that an external field can cause a breakdown of ergodicity and stabilize a pair of magnetization plateaus: metastable states that interrupt progress towards the ground state. 
These plateaus are extremely stable despite the absence of frustration or intrinsic disorder (the conventional causes of freezing behavior). 
We described the plateaus in terms of local spin configurations that are stable under our dynamics. 
The plateaus consist of an ensemble of states where each spin is at the center of one of these stable local configurations. 
From such states, all single spin flips increase the energy. 
This corresponds to an extremely rough energy landscape without a clear gradient pointing towards the ground state (at least from the perspective of these dynamics). 
The local energy minima are so numerous that the odds of reaching the true ground state are vanishingly small, since paths to the ground state will almost always intersect one of the plateau states, where the simulation will become stuck. 
The underlying mechanism for this behavior is the field lifting the degeneracy of local spin configurations, eliminating reversible zero-energy Monte Carlo updates that are crucial for finding the correct ground state. 

The process of finding the ground state of the Ising model from a random initial state is related to a broad class of optimization and gradient descent problems in fields such as machine learning. 
Often, the energy landscape is described in terms of a simple height function in some high-dimensional space with local minima that look like valleys. 
This description can be misleading: although the energy landscape is indeed a surface in some high-dimensional space, dynamics often include nonlocal moves, which means that the choice of dynamics can dramatically alter the notion of what other states are `nearby' and therefore of what states appear to be local energy minima \textit{to the optimization algorithm}. 
With our single spin flip dynamics, each state is connected to exactly $N$ other states.\footnote{(all the states that can be reached by a single spin flip)} 
Under Kawasaki dynamics (where pairs of antiparallel nearest-neighbor spins are flipped \cite{kawasaki1966,tartaglia2018}), each spin state would be connected to a \textit{totally different} set of `nearby' states and the \textit{apparent} local energy minima (with respect to the dynamics) would therefore be different as well. 
Indeed the field does not affect the Kawasaki transition probabilities at all. 
From these two examples (Metropolis and Kawasaki), we can see that the notion of which states are nearby, and therefore which states appear to be local energy minima, are completely dependent on the choice of dynamics. 
By analogy, gradients also depend on the dynamics: a local energy minimum under one choice of dynamics might lie on a steep slope under another. 

Although the details of the magnetization plateaus depend on the specific update scheme, the underlying principle causing the breakdown of ergodicity is quite general: 
a Markov chain can become non-ergodic when there are few \textit{reversible} updates available.\footnote{Here, reversible means updates that do not dramatically change the energy.}  
Naively, the fastest way to the ground state is to use updates with a large negative $\Delta E$, and by that logic, $\Delta E=0$ updates are a waste of time. 
In fact, $\Delta E=0$ updates are critical for avoiding local energy minima because they allow movement along equal-energy paths to find the true global minimum. 
This fact is implicitly incorporated into many Monte Carlo update schemes (like cluster methods) that try to find groups of spins, etc., that can be updated without changing the energy. 

Here we have made no attempt to `fix' the frozen dynamics. 
Using different dynamics or simulated annealing might allow the system to reach the ground state, but our goal was to study the freezing process itself. 
Understanding how Monte Carlo methods fail is crucial because MCMC methods are depended upon to serve as a reliable, unbiased `numerical experiments' with well-defined statistical error and few approximations.\footnote{In contrast to various numerical techniques based on perturbation theory or other expansions, which often include some poorly-defined systematic error.} 
In the case we have studied here (focusing on the strongly-frozen plateau regime), the dynamics are slowed to a complete halt after a very brief transient. 
The MC sweeps rapidly flip all the `available' spin flips; once every spin is at the center of one of the stable local states, no further changes are possible. 
Ironically, this makes this particular system very easy to study since it freezes so quickly and so completely that there is no need for long simulations, but it may still yield useful comparisons to freezing in systems with intrinsic disorder such as the random field Ising model \cite{belanger1991} or spin glasses \cite{binder1986}, which are more challenging to study directly.

Our findings suggest a number of interesting avenues for further research. 
The type of plateaus that occur in the 2D Ising AFM are likely to be quite general, and similar plateaus probably occur with other lattices, Hamiltonians and dynamics; examining these broader applications could uncover universal features. 
Using the framework described in \cref{s:conf}, we can make some immediate predictions about quenches in the one-dimensional (1D) and three-dimensional (3D) Ising antiferromagnets. 
In 1D, our brief tests show evidence for a single magnetization plateau from $0<h<2$ around \mbox{$m\approx 0.14$} (where $h=2$ is the saturation field). 
The freezing mechanism is the same: the field stabilizes domain walls with an excess $+$ spin: $...-+-++-+-...$, but we have not investigated this case in detail. 
In 3D, we expect three plateaus.  
There are six nearest neighbors, so $y_{\rm 3D} = 0, \pm2, \pm4, \pm6$, and therefore we expect $\Delta E=0$ updates for $h=0,\pm2,\pm4,\pm6$, with plateaus between those points. 

It may be possible to develop analytical approaches to derive magnetizations and configuration populations in the $m_1$ and $m_2$ plateaus from first principles. 
One method may be to enumerate all possible plateau states based on the domain wall rules we identified in \cref{s:plateau} using a scheme similar to Ref.~\onlinecite{muller1977}. 
That enumeration could quantify the scaling of the number of plateau states and possibly allow an analytical derivation of quantities such as the magnetization. 
Alternatively, one could attempt to identify a connection to percolation theory that describes the plateaus as Ref.~\onlinecite{barros2009} did for the striped states in the ferromagnet. 
Even if such approaches remain elusive, there is still much to be learned from a more detailed study of the plateau states themselves. 
For example, the freezing halts the coarsening process, but it is not immediately clear if the distribution of domain sizes in the plateau states corresponds directly to a point along the conventional Ising coarsening process \cite{lopes2020}. 
It also might be instructive to investigate non-square aspect ratios,\footnote{Here the aspect ratio is $\left(\frac{\rm height}{\rm width} \right)$.} which could affect the stability or magnetization of the plateaus. 
For the ferromagnet, the aspect ratio affects the probability of becoming stuck in the striped state \cite{barros2009}. 

Finally, we have focused on the strongly-frozen regime near the central flat portion of the plateaus where the relaxation time is effectively infinite. 
There are a number of interesting questions at nonzero temperature, for example: what is the maximum temperature $T^*$ at which the plateaus appear? 
The edges of the plateaus and valleys of ergodicity may yield still richer physics. 
In this regime, the relaxation time is finite (but long) and there are nontrivial finite size and temperature effects.

\section*{Acknowledgements}

I would like to thank my friends and colleagues Anders W. Sandvik, Ying-Jer Kao, Paul L. Krapivsky and Wen-Han Kao for their advice and feedback, and my wife, Vanessa Calaban, for her unwavering support (and proofreading). 
The code I have used here \cite{afmIsingCode} is based on a program originally written by Anders Sandvik with my own additions and modifications. 

\appendix

\section{Finite-size scaling\label{s:fss}}

\begin{figure} 
\includegraphics[width=80mm]{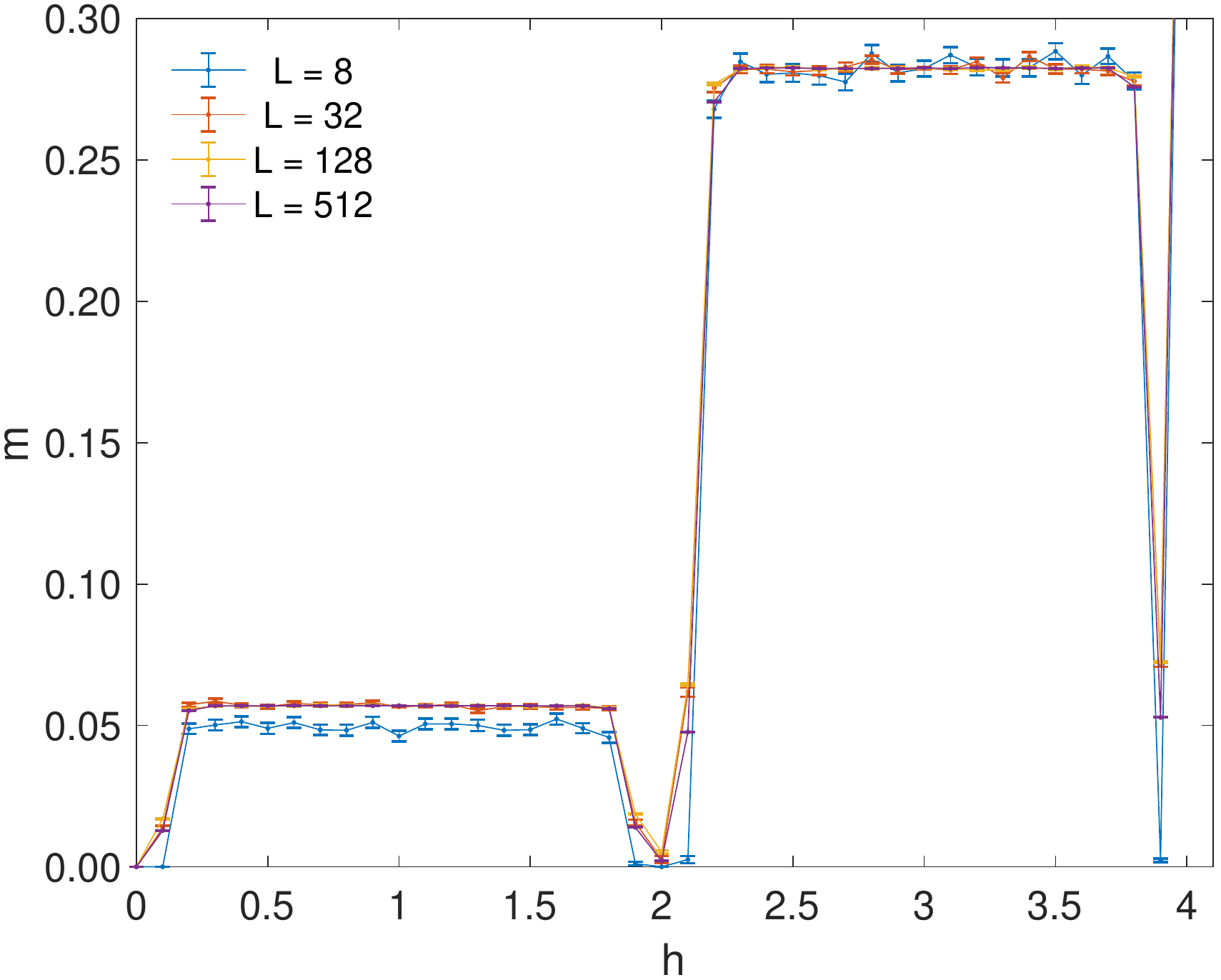}
\caption{Finite-size scaling of magnetization after quenches to $\beta=32$ with periodic boundary conditions. Each point is an average over 100+ independent quenches. The finite-size effects vanish rapidly (compare to \cref{f:magcurve}).  \label{f:fssMag}}
\end{figure}

One of the remarkable things about the metastable magnetization plateaus we describe here is that the finite-size effects vanish so rapidly.\footnote{Although we use large system sizes in our analysis ($L=512$ in \cref{f:magcurve}), these are much larger than is necessary.}
This suggests that when the system becomes stuck, the correlation length is still small. 
In \cref{f:fssMag} we show the magnetization curves for a quench to $\beta=32$ for several sizes $L=2^3,2^5,2^7,2^9$ (compare to \cref{f:magcurve}). 
For the $m_1$ plateau, there are some finite-size effects for $L=8$, largely due to the proximity of the plateau states to the ground state.\footnote{For an $8\times8$ system $m=0.0569$ corresponds to just three or four excess $(+)$ spins.} 
For larger systems, the $m_1$ plateau appears fully converged to the thermodynamic limit. 
In the $m_2$ plateau, the finite-size effects are weaker and even the $8\times8$ system agrees with the largest size within error bars. 
There are more prominent finite-size effects at higher temperatures and especially around the valleys of ergodicity, but we do not discuss those regimes here. 

\begin{figure} 
\includegraphics[width=80mm]{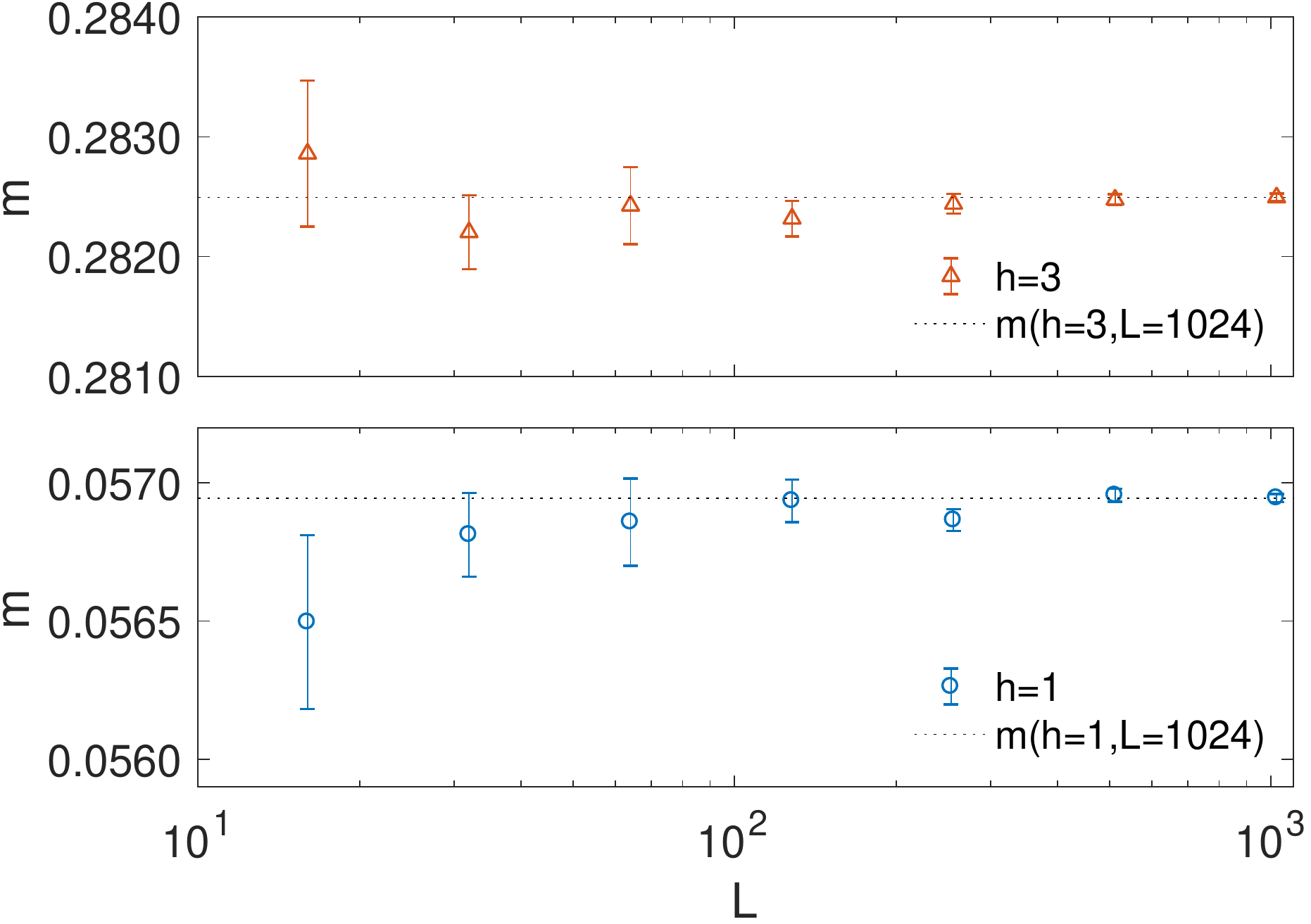}
\caption{Finite-size scaling of magnetization for (\textbf{bottom}, $h=1$) $m_1$ and (\textbf{top}, $h=3$) $m_2$ plateaus for quenches to zero temperature with periodic boundary conditions, (semilog scale). Each point is an average over many independent quenches (at least 400, more for smaller systems). Error bars are smaller for larger systems due to spatial self-averaging. \label{f:fssPlat}}
\end{figure}

In \cref{f:fssPlat} we show the finite-size scaling at $T\to 0$ in the center of each plateau (note: the $x$-axis is a log scale). 
The error bars in all cases are very small, but even so, $L=16$ is barely distinguishable from $L=1024$.

\section{Open boundary conditions\label{s:obc}}

\begin{figure} 
\includegraphics[width=80mm]{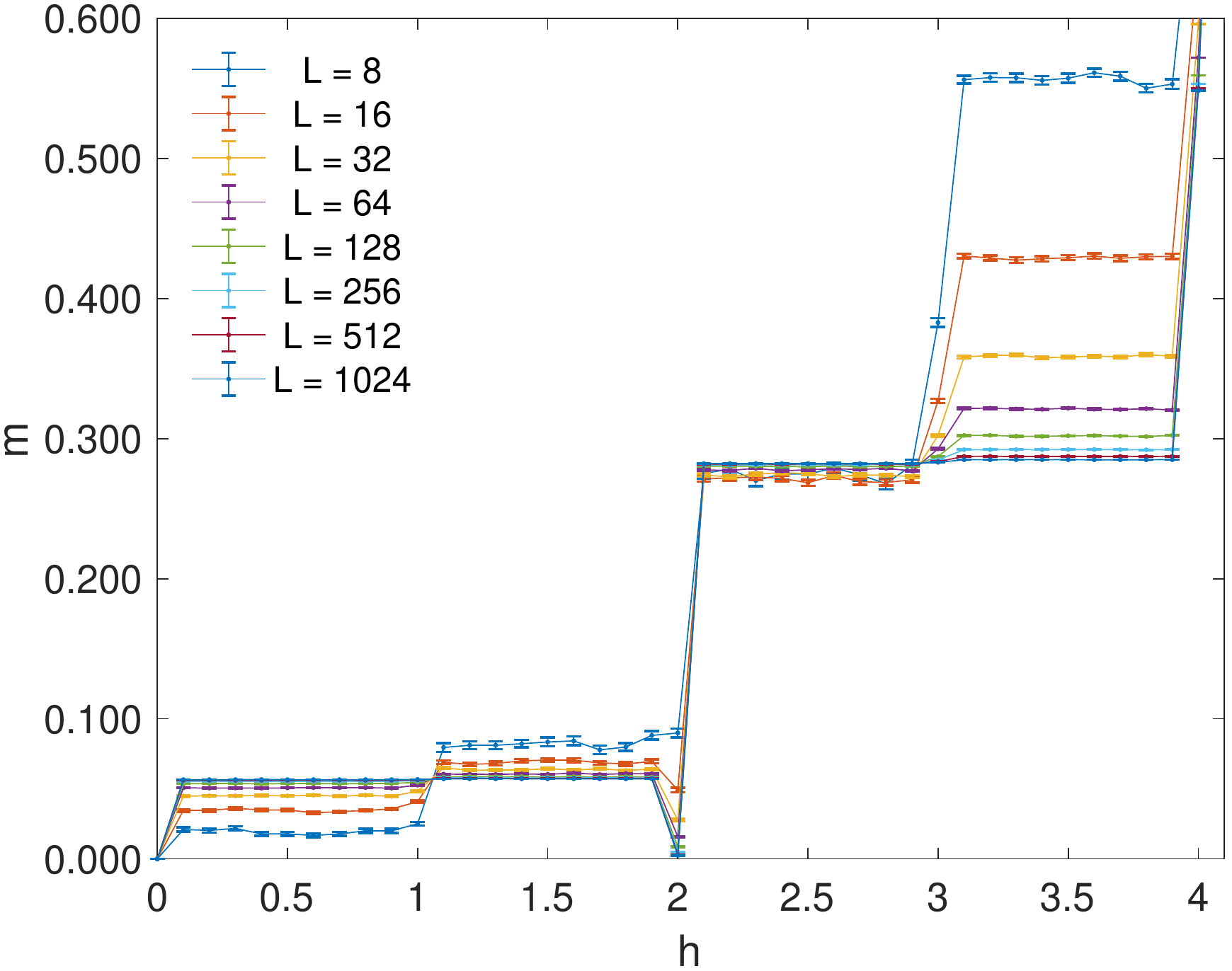}
\caption{Finite-size scaling of magnetization after quenches to $\beta=128$ with open boundary conditions. For OBC, the finite-size effects are much more pronounced (compare to \cref{f:magcurve,f:fssMag}). Each point is an average over $\approx200$ independent quenches. \label{f:fssOBC}}
\end{figure}

In the main text we have focused almost entirely on the case of periodic boundary conditions (PBC), but it is worth comparing to open boundary conditions (OBC). 
The behavior is largely the same, but for OBC there are prominent finite-size effects that can be described in terms of the local-configuration framework developed in \cref{s:conf}. 
The bulk spin states appear to be the same as for PBC, but the edges have different stable local configurations so they have a different average magnetization. 

Along the boundaries the spins have only three neighbors, so the possible values of $y_{\rm edge} = \pm1 \pm3$. 
There are new fields where the stable local configurations change $(h=\pm1,\pm3)$. 
The effect of the edge spins is to break each plateau up into two subplateaus: $0<h<1$, $1<h<2$ and $2<h<3$, $3<h<4$, which recombine in the thermodynamic limit. 
We see these subplateaus and their finite-size scaling in \cref{f:fssOBC} (compare to PBC in \cref{f:fssMag}). 
In principle, it might be possible that the $\Delta E=0$ updates along the edges at $h=\pm1,\pm3$ could cause new valleys of ergodicity, but we see no signs of this actually occurring (at least not at such low temperature). 

The corner spins have only two neighbors, which can add up to $y_{\rm corner} = 0, \pm2$. 
The bulk spins already have reversible spins flips at $h=0,\pm2$, so the corner spins do not contribute anything new. 

\begin{figure} 
\includegraphics[width=80mm]{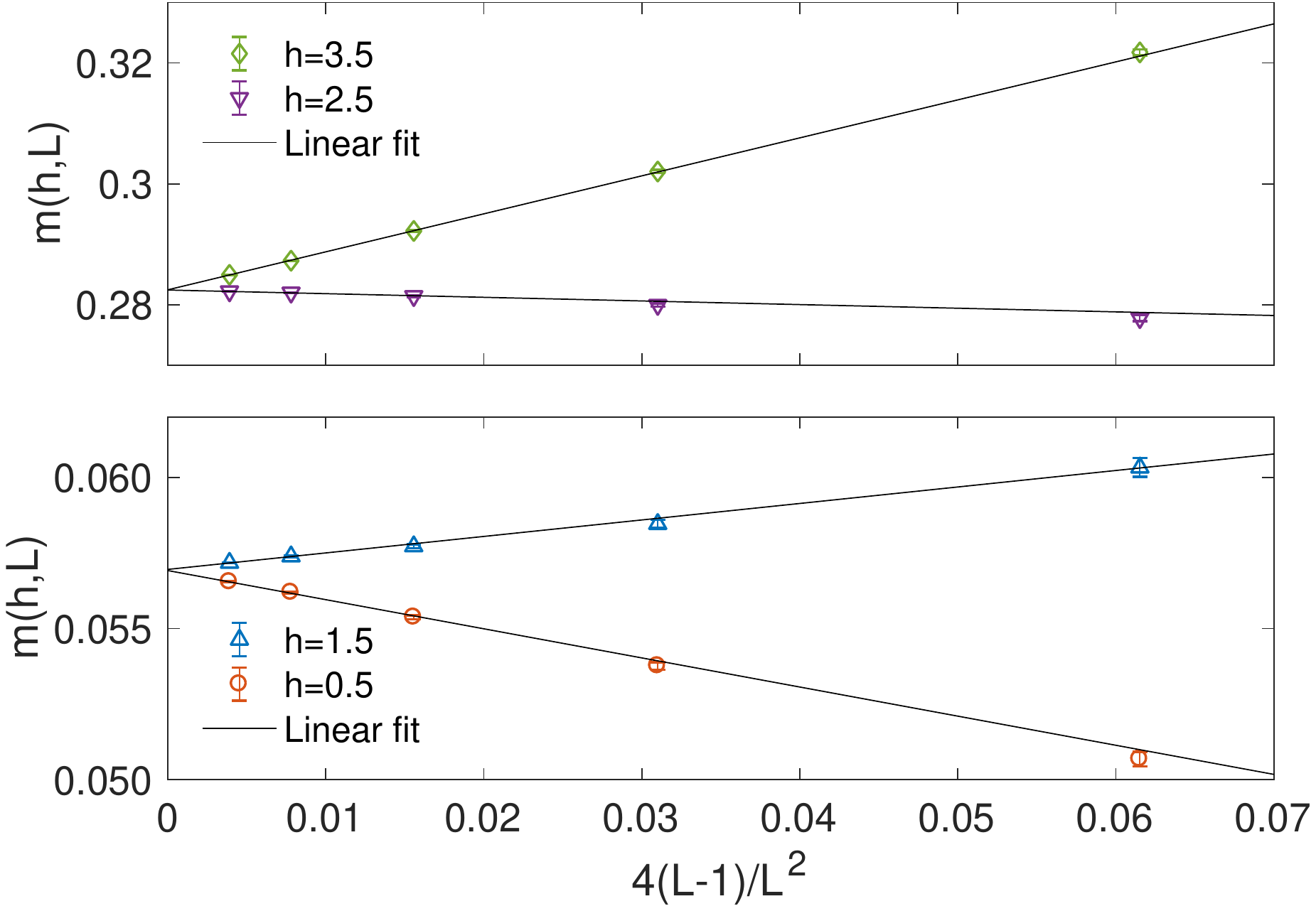}
\caption{Finite-size scaling of the plateau magnetization with open boundary conditions under quenches to $\beta=128$ as a function of $n_{\rm edge}/L^2$. (\textbf{bottom}) Both halves of the $m_1$ plateau: \mbox{$h=0.5,1.5$}. (\textbf{top}) Both halves of the $m_2$ plateau with \mbox{$h=2.5,3.5$}. Each point is an average over 200 independent quenches.  \label{f:platOBC}}
\end{figure}

The most notable consequence of OBC is strong finite-size effects with well-defined scaling. 
As $L$ grows, the contribution from the edge states shrinks.
In \cref{f:platOBC} we examine the finite-size scaling of the magnetization in the new subplateaus. 
As expected, the boundary conditions become irrelevant in the thermodynamic limit.
The finite-size deviation is linear in $n_{\rm edge}/L^2$, where $n_{\rm edge} = 4(L-1)$ is the number of sites along the boundaries of the $L\times L$ square lattice. 
For $0<h<1$ and $1<h<2$, the magnetization converges to the PBC value of $m_1$ as $L\to \infty$. 
For $2<h<3$ and $3<h<4$, the magnetization converges to the PBC value of $m_2$. 
Plotting $m(h,L)$ against $4L^{-2}(L-1)$ we recover an excellent linear scaling law.

\section{Local configurations at $T=\infty$ \label{s:tinf}}

For $T=\infty$, each spin will independently take values $\sigma_i=\pm1$ with equal probability. 
There are $2^5=32$ possible states of the center spin and its four neighbors. 
Of those $2^4=16$ have the center spin up, and $2^4$ have the center spin down.  
For the operators $C^{-4}_{-1}$, $C^{-4}_{+1}$, $C^{+4}_{-1}$ and $C^{+4}_{+1}$, there is only one way to arrange four parallel neighbors, so those each appear with probability $P=(\frac{1}{2})^5=3.125\%$. 
For configurations with three parallel and one antiparallel neighbors $C^{-2}_{-1}$, $C^{-2}_{+1}$, $C^{+2}_{-1}$ and $C^{+2}_{+1}$, there are ${ 4 \choose 1}=4$ states for each; therefore, those states each appear with probability $P=4(\frac{1}{2})^5=12.5\%$. 
Finally for the configuration with two $(+)$ and two $(-)$ neighbors ($C^{0}_{-1}$ and $C^{0}_{+1}$), there are ${4 \choose 2}=6$ possible configurations for each and those states each appear with probability $P = 6(\frac{1}{2})^5 = 18.75\%$. 
These predictions are confirmed by numerical tests on 7,000 random $64\times 64$ spin configurations in \cref{tab:pop}.

\section{Mapping onto other problems at $h_s$\label{s:hardsquares}}

At the saturation point $(h=h_s)$, the AFM ground states and the fully-polarized state have the exact same energy, and all three can coexist at no energy cost. 
At zero temperature the ground state jumps from \mbox{$m(h=4-\epsilon)=0$} to \mbox{$m(h=4+\epsilon)=1$}, which is smoothed out at finite temperature. 
The saturation point has connections to two other statistical physics problems. 
The first is the reversible random sequential adsorption process \cite[p. 220]{krapivsky2010} where the `empty' state is the fully-polarized, all-$C^{+4}_{+1}$ configuration, and $(-)$ spins (or $C^{+4}_{-1}$ objects) randomly adsorbed onto sites.
The restriction of local spin configurations becomes a rule that no $(-)$ spin will adsorb onto a site where any of its nearest neighbors are $(-)$. 
The second is the so-called hard squares problem \cite{huckaby1984,fernandes2007,liu2000,hu1989} with $\mu=0$, where hard-core particles with a radius of one (excluding nearest neighbor sites) are adsorbing on a lattice.

\bibliography{bibstuff}

\end{document}